# Web Based Information System for Heat Supply Monitoring

**Borislav Stoyanov and Strahil Strahilov**

Department of Computer Informatics
Faculty of Mathematics and Informatics
Konstantin Preslavski University of Shumen, 9712 Shumen, Bulgaria

**Abstract**
*The paper presents web based information system for heat supply monitoring. The proposed model and information system for gathering data from heating station heat-flow meters and regulators is software realized. The novel system with proved functionality can be commercialized at the cost of minimal investments, finding wildly use on Bulgarian market as cheap and quality alternative of the western systems.*

**Subject Codes**: 68M11, 68U35, 97R50.

**Keywords**: heat, supply, monitoring, control, information system.

## 1   Introduction

The introduction of dispatching systems for monitoring and control of heating systems is a relatively new trend, which is quickly imposed with the increase of energy efficiency requirements, associated with the increased prices of fuel and energy and the growing concern for environmental protection. This type of systems are actually a variety of the well-known from the industry SCADA (Supervisory Control And Data Acquisition Systems) systems [1]. The difference is that instead of PLC (programmable logic controllers), used in industrial systems, the management of heating in buildings is realized with specialized digital controllers for heating installations management. Heat-flow meters are usually included in the system. Communication with the dispatch (operator) station can be realized in different ways. This is the main feature which differentiates those systems.

The local communication with the heat-flow meters is done almost always by standard Message Bus (Mbus) protocol. Based on this, the objective of this article is to develop a web based control and information system for data collection from heating installations heat-flow meters and regulators. To achieve this goal in the following sections of this article the following three main tasks are solved:

1. Analyzing the types of dispatching systems for control and gathering information from buildings heating installations.

2. Proposing conception of web based control and information system for heat sypply monitoring.

3. Software realization of the web based heating installation control and information system model.

## 2 Analysis of the types of dispatching systems for building heat stations control and gathering information

**DIAL UP BASED DISPATCHING SYSTEMS**

These dispatching systems are based on modems that use relatively old technology, based on the so-called SMART modems supporting a standard set of AT commands. A connection with channel commutation is build, where the maximum speed of data transmission is relatively low 56 kbits/s. An advantage of this technology is the low cost of the whole product.

**WIRELESS BASED DISPATCHING SYSTEMS**

The systems based on wireless communication, may conditionally be classified into two types:

*GPRS systems* using standard mobile phone service – the modern way of transmitting data where communication is done by using standard Internet protocols.

Irreplaceable in hard to reach places where no Internet and PSTN (Public Switch Telephone Network) web infrastructure is developed.

*Systems based on radio communication* – specialised radio stations are used to transmit data on different frequency channels. Relatively old, low-budget technology, using low speed flows in the information transmission. The block diagram of a system based on radio communication is analogous to GPRS using radio stations, operating in particular frequency range. All available frequencies must be licensed, which limits the application of such systems.

**WEB BASED DISPATCHING SYSTEMS**

These systems use mostly the existing Internet (Broadband Internet) in the building – cable or ADSL. ADSL is kind of broadband Internet access. Works by splitting the existing telephone line signal into two - one for voice services, and the other for data. Thus, telephone and Internet services can be provided over the same line and at the same time. This asymmetric layout (with much higher speeds towards the user compared to the opposite direction) is ideal for those who want to use the Internet to surf the web pages (downloading of web pages from Internet server) or receiving e-mail (downloading electronic messages from the mail server). Communication is performed by standard Internet protocols such as TCP/IP and HTTP. They are characterized by high speed, the ability to transmit large volumes of data, simplified software implementation, standard communication protocols.

Generalized classification of control and information systems is shown in **Table 1**.

**Table 1** Classification of the dispatching systems

| Types of systems | Advantages | Disadvantages |
|---|---|---|
| **1**. Modem (dial up) communication. | - relatively cheap and accessible way of communication | - old technology<br>- low speed |
| **2.** Systems based on radio communication. | - relatively cheap and accessible way of communication | - old technology<br>- low speed |
| **3.** GPRS systems using standard mobile network operator. | - standard network of GPRS operator<br>- possibility of communication in the inaccessible places without Internet | - relatively expensive<br>- harder for software implementation |
| **4.** Systems based on web communication. | - high speed<br>- capability of transmission with high capacity data<br>- easy to realize the software implementation<br>- standart communication protocols<br>- relatively cheap and accessible way of communication | - need for Broadband Internet |

On the basis of **Table 1** the most effective are the web based systems as in the absence of an Internet connection a GPRS oriented system can be build.

## 3 Conception of the web based control and information system of the heat supply monitoring

The functionality of the system includes the collection of data from remote devices and record in a databases. The devices can be both metering equipment (electrometers and heat-flow meters) and controllers (regulators via RS485) for intelligent management of heating systems. It is possible to transfer data in both directions, i.e. not only collecting data, but sending command instructions to the regulators (mode of operation change, settings change, etc.).

The system uses standard protocols, implemented in GPRS/GSM modems for data transportation in Internet over a wireless connection. The local connection between the modems and reading devices is through standard interfaces like Mbus for measuring devices or specific communication protocols regulators reading.

For the system realization are used the following protocols and programming languages: HTTP, PHP [3], HTML, JavaScript, MySQL [2], .NET, C#.

Some closely related information system examples exist in the papers [4-7].

# 4 Software realization of the web based control and information system for heat supply monitoring

The software module represents a Windows Forms application implemented on C# .NET. The user interface allows you to simulate the values of eight sensors and the conditions of eight actuators (pumps), whose momentary conditions are able to be transmitted to the server at user request. This is done by activating a GET HTTP request to the URL of the zapis_danni.php script. The data is transmitted as a text string in the text field of the GET request.

The created simulation program has two main classes: Program and Form1. These two classes are implemented in namespace test01.

Class Program – this method is the entry point of the application, interpreting in its structure the method Main(). This method sets the style of the application, the text format and defines the way of communication between the classes Program и Form1.

Class Form1 – in this class the main methods giving functiontality of the application are realized:

The method backgroundWorker1_DoWork(object, System.ComponentModel.DoWorkEventArgs) sets a double-stranded mode of operation to the applied software. It defines variables that realize communication with the HTTP server (query, response).

The method button1_Click(object, System.EventArgs) defines all the needed variables with the values entered by the user; they are recorded in the appropriate format (string) and sent to the server, where they are written in a text file.

The method button2_Click(object, System.EventArgs) sends information to the HTTP server for the mode of operation entered by the user.

The methods from checkBox1_CheckedChanged(object, System.EventArgs) to checkBox8_CheckedChanged(object, System.EventArgs) define the control buttons on the outgoung mechanisms (pumps) and the values that they can accept (on and off).

The method Dispose(bool) – releases the resourses on the dialog window

The method Form1() – in it the necessary variables are initialized; sets the range of the preset sensors (temperatures); defining the variables that characterize the actuators; setting asynchronous double-stranded mode of operation; setting a variable tracking the mode of operation;

The method Form1_Exit(object, System.EventArgs) – releases the resourse of the double-stranded mode.

The method InitializeComponent() – visualisation of variables.

The methods from label16_Click(object, System.EventArgs) to label22_Click(object, System.EventArgs) – puts variable labels.

The method SetText(string, int) monitors the control mechanisms condition information in the periodic communication between the server and the application

The methods from trackBar1_Scroll(object, System.EventArgs) to trackBar8_Scroll_1(object, System.EventArgs) – define the value settings of the sensors/temperatures.

The server used is Apache HTTP server with configured PHP module. Two scripts are implemented for gathering and visualiation of the data:

zapis_danni.php – script for gathering data, that creates and fills a table with data from the simulation application.

danni.php – script for visualisation of the database data.

## 5 CONCLUSION

The analysis of the dispatching systems shows a priority in choosing systems that use Internet connection. The proposed model of a web based control and information system for gathering data from heating station heat-flow meters and regulators is software realized. The novel system with proved functionality can be commercialized at the cost of minimal investments, finding wildly use on Bulgarian market as cheap and quality alternative of the western systems.

## 6 GUIDELINES FOR FURTHER DEVELOPMENT

The guidelines for further development can be summarized in the following areas:
- Application of remote control and monitoring of sensors parameters in the area of heat supply, industry and energetics;
- Implementing of the web based system for the needs of the energy monitoring of a eco company buildings;
- Increasing the possibilities of the web based system for remote measuring of electrometers and heat-flow meters.